\documentstyle[prl,aps,twocolumn,epsf,epsfig]{revtex}
\begin{document}
\twocolumn[\hsize\textwidth\columnwidth\hsize\csname@twocolumnfalse\endcsname
\draft

\title{Magnetic Phase Diagram of Ca$_{2-x}$Sr$_x$RuO$_4$ Governed by
  Structural Distortions}

\author{Z. Fang$^1$ and K. Terakura$^{2,3}$}

\address{ $^1$JRCAT, Angstrom Technology Partnership, 1-1-4 Higashi,
   Tsukuba, Ibaraki 305-0046, Japan}

\address{ $^2$JRCAT, National Institute for Advanced Interdisciplinary
   Research, 1-1-4 Higashi, Tsukuba, Ibaraki 305-8562, Japan}

\address{ $^3$ Tsukuba Advanced Computing Center, 1-1-4 Higashi,
   Tsukuba, Ibaraki 305-8561, Japan}

\date{\today }
\maketitle

\begin{abstract}
  We constructed, by the first-principles calculations, a magnetic
  phase diagram of Sr$_{2}$RuO$_4$ in the space spanned by structural
  distortions. Our phase diagram can qualitatively explain the
  experimental one for Ca$_{2-x}$Sr$_x$RuO$_4$. We found that the
  rotation and the tilting of RuO$_6$ octahedron are responsible for
  the ferro- and antiferro-magnetism, respectively, while the
  flattening of RuO$_6$ is the key factor to stabilize those magnetic
  ground states.  Our results imply that the magnetic and the
  structural instabilities in Sr$_2$RuO$_4$ are closely correlated
  cooperatively rather than competitively.
\end{abstract}

\pacs{PACS numbers: 75.30.Kz, 74.70.-b, 71.27.+a}
]

\narrowtext

Both the magnetic and the structural instabilities are essential
issues for the unconventional superconductivity in
Sr$_2$RuO$_4$~\cite{nature,NMR,Impurity,time_reversal,flux,Andreev},
which is the only example of a noncuprate layered perovskite
superconductor.  It was first suggested that the Sr$_2$RuO$_4$ is
close to the ferromagnetic (FM) instability~\cite{p-wave} with strong
FM spin fluctuations, which may naturally lead to a spin-triplet
$p$-wave pairing mechanism~\cite{p-wave,Mazin1,Tewordt,p-wave2}.
However, the recent observation~\cite{Sidis} of sizable
antiferromagnetic (AF) incommensurate spin fluctuation, due to the
Fermi surface nesting~\cite{Mazin2}, indicates that more careful
studies are needed. As for the structural aspect, it was pointed out
by experiments that Sr$_2$RuO$_4$ is very close to the structural
instability with respect to the RuO$_6$ rotation~\cite{phonon}.  With
such a situation, one may consider that three kinds of instabilities,
superconducting, magnetic and structural ones, may compete.
Nevertheless, the correlation among those instabilities has not been
fully discussed.  It was found recently~\cite{plummer} that the
cleaved surface of this material is reconstructed to form the c(2x2)
structure which can be regarded as the frozen RuO$_6$ rotation
mentioned above. Furthermore, the density-functional calculation
predicts that the surface ferromagnetism is strongly stabilized by the
structural reconstruction~\cite{plummer}.  This prediction suggests
that the structural and magnetic instabilities cooperate rather than
compete, although the surface ferromagnetism has not been
experimentally confirmed up to now.

On the other hand, the recent studies on the Ca$_{2-x}$Sr$_x$RuO$_4$
suggest strongly the cooperative feature of the structural and
magnetic instabilities in the bulk.  Moreover, the system shows a very
rich phase diagram and provides us with an opportunity to analyze the
correlation between the magnetism and the structure more extensively.
Below is a brief description of the experimental observation for
Ca$_{2-x}$Sr$_x$RuO$_4$ by Nakatsuji {\it et al}.~\cite{Nakatsuji}.
With the Ca substitution for Sr, the system is successively driven
from the non-magnetic (NM) 2-dimensional (2D) Fermi liquid
($x\sim2.0$) to a nearly FM metal ($x\sim0.5$), an
antiferromagnetically correlated metal ($0.2<x<0.5$), and finally an
AF insulator ($x<0.2$).  Since the substitution is isovalent, the
dominant effects are the structural modifications due to the reduced
ionic size of Ca compared with Sr.  Evidence~\cite{Friedt} has been
presented by neutron scattering that the structural distortions caused
by the Ca substitution correlate with the changes in the magnetic and
the electronic properties.

The main aim of this letter is to study how and why the magnetism of
Ca$_{2-x}$Sr$_x$RuO$_4$ is affected by structural distortions.  In
order to extract essential aspects, we assume that for a given crystal
structure, the electronic structure is not affected by the relative
content of Ca and Sr.  Therefore, in the following, we study the
stable magnetic phases of Sr$_2$RuO$_4$ for given structural
distortions.  Three types of structure distortions, i.e. RuO$_6$
octahedron {\it rotation} about the $c$-axis, RuO$_6$ {\it tilting}
around an axis parallel to the edge of octahedron basal plane and the
{\it flattening} of RuO$_6$ along the $c$-axis are identified from
experiments~\cite{Friedt}. Our phase diagram can qualitatively explain
the experimental phase diagram of Ca$_{2-x}$Sr$_x$RuO$_4$,
demonstrating the crucial roles of structural distortions for the
tuning of electronic and magnetic properties, and further supporting
our previous prediction for the surface . In particular, we found that
the RuO$_6$ rotation can enhance the FM instability significantly,
while the combination of tilting and rotation of RuO$_6$ is
responsible for the enhancement of AF instability. Furthermore, we
point out that the flattening of RuO$_6$ is a key factor to stabilize
the magnetic (both FM and AF) ground states.  The basic physics
governing the phase diagram can be understood in terms of the strong
coupling between the lattice and the magnetism through the orbital
degrees of freedom. Our results strongly suggest that, in
Sr$_2$RuO$_4$, the magnetic fluctuations can be significantly enhanced
by the structural fluctuations, implying the necessity of
reconsidering the coupling mechanism in the bulk superconductivity.

The calculations were performed with the first-principles plane-wave
basis pseudopotential method based on the local density approximation
(LDA). The validity of LDA treatment for ruthenates was demonstrated
in Ref.~\cite{Mazin1,Mazin3}. The $2p$ states of oxygen and $4d$
states of Ru are treated by the Vanderbilt ultrasoft
pseudopotential~\cite{PP}, while the norm-conserving scheme~\cite{NC}
is used for other states. The cutoff energy for the wave function
expansion is 30~Ry. The $k$-point sampling of the Brillouin zone (BZ)
was well checked to provide enough precision in the calculated total
energies.  The theoretically optimized lattice parameters
$a=3.84$~\AA~and $c=12.70$~\AA~ for the bulk Sr$_2$RuO$_4$ are in good
agreement with the experimental ones $a=3.86$~\AA~ and
$c=12.73$~\AA. The degree of flattening of RuO$_6$ octahedron $\lambda
$ is defined by $\lambda = d_c / d_{ab}$ with $d_c$ ($d_{ab}$)
denoting the Ru-O bond length along the $c$-axis (in the $ab$-plane)
with the RuO$_6$ volume fixed.  Rotation and tilting of RuO$_6$
octahedron are operated with the Ru-O bond lengths fixed.  In order to
construct the magnetic phase diagram, the lowest energy magnetic phase
for each crystal structure is searched for among different (NM, FM and
AF) phases.  In the present work, we focus our attention only on
phases described within the c(2x2) unit cell.  The soft phonon mode of
$\Sigma _3$ at the zone boundary in Sr$_2$RuO$_4$~\cite{phonon} and
the AF state of Ca$_2$RuO$_4$~\cite{CRO} are in this category.

Figure 1 shows the calculated phase diagram~\cite{comment1}, while the
Table 1 summarizes the calculated total energies and magnetic moments
for some particular points in the phase diagram.  Hereafter, $\phi$
and $\theta$ denote the rotation angle and the tilting angle,
respectively.  The apical oxygen and the oxygen in the $ab$-plane are
called O(2) and O(1) respectively. From right to left of the phase
diagram, first the RuO$_6$ starts to rotate along the $c$-axis by up
to 12$^{\circ}$, and then with the 12$^{\circ}$ rotation being fixed,
the RuO$_6$ starts to tilt up to 12$^{\circ}$.  The structural
analysis by the neutron scattering~\cite{Friedt} allows us to make a
one-to-one correspondence between the structural changes, i.e. the
horizontal axis of our phase diagram, and the doping level $x$ in
Ca$_{2-x}$Sr$_x$RuO$_4$. For $x=2.0$ (Sr$_2$RuO$_4$), the system has
$I4/mmm$ symmetry with $\phi=\theta=0^\circ$, corresponding to the
right end of our phase diagram. With reduction of $x$, RuO$_6$ starts
to rotate and the symmetry is reduced to $I4_1/acd$ until $x=0.5$
(Ca$_{1.5}$Sr$_{0.5}$RuO$_4$), where $\phi=12.78^\circ$ and
$\theta=0^\circ$ at 10 K. With further reduction of $x$, RuO$_6$
starts to tilt and the symmetry is further reduced to $Pbca$ until
$x=0.0$ (Ca$_2$RuO$_4$) where $\phi=11.93^\circ$ and
$\theta\sim12^\circ$ at low temperature, corresponding to the left end
of our phase diagram. It was also pointed out by the
experiment~\cite{Friedt} that, from $x=2.0$ to $x=0.5$, the degree of
flattening $\lambda$ remains almost constant ($\sim 1.07$), while from
$x=0.5$ to $x=0.0$, the rotation angle $\phi$ is almost unchanged
($\sim 12^\circ$). Three representative experimental points are shown
in our phase diagram by triangles.  Now, the basic tendency suggested
by our phase diagram is that the RuO$_6$ rotation will drive the
system from a NM state to a FM state, while the subsequent tilting
plus the flattening of RuO$_6$ will push the system to an AF region.
This general tendency is quite consistent with the experimental
results.  It should be noted that the rich experimental phase diagram
can be simply understood in terms of the close coupling between
structural distortions and magnetism. Although the real long range FM
ordering in Ca$_{1.5}$Sr$_{0.5}$RuO$_4$ is not confirmed yet, the
significant enhancement of spin susceptibility in this doping level
undoubtedly implies the strengthening of FM instability.  Another
important aspect in our phase diagram is that the flattening of
RuO$_6$ is so important not only for the AF state but also for the FM
state. This suggests that {\it simply by uniaxial pressure, the
  Sr$_2$RuO$_4$ can be driven from the NM state to a FM state}.

The basic questions concerning our phase diagram are: 1) Why are the
RuO$_6$ rotation and tilting correlated with the tendency to the FM
and AF states? 2) Why is the RuO$_6$ flattening so important for the
magnetic solutions?  Before answering these questions, let us discuss
the role of each $4d$ orbital in the electronic properties of
Sr$_2$RuO$_4$, which is essential to our later discussions.  The three
Ru $t_{2g}$ orbitals ($d_{xy}$, $d_{yz}$, $d_{zx}$) hybridize with
each other only very weakly in tetragonal Sr$_2$RuO$_4$.  Therefore
each orbital plays distinct roles.  The projected density of states
(DOS) shown in Fig.2(a) indicates that the $d_{xy}$ orbital
contributes dominantly to the well-known van Hove singularity (VHS)
just above the Fermi level.  The $\gamma$ Fermi surface has the
character of $d_{xy}$.  It is mostly responsible to FM spin
fluctuation due to the high DOS around the Fermi level.  On the other
hand, $d_{yz}$ and $d_{zx}$ orbitals contribute to the $\alpha$ and
$\beta$ Fermi surfaces and produce the incommensurate spin fluctuation
coming from the strong nesting effect due to the quasi-one-dimensional
nature of those states.  The calculated bare spin susceptibility shown
in Fig. 3(a) for the undistorted compound has the incommensurate peak
at {\bf Q}=($2\pi/3a, 2\pi/3a$), being consistent with previous
calculations~\cite{Mazin2}. In real materials, the two factors,
i.e. the FM instability due to the high DOS at the Fermi level
combined with the $q$ dependent Stoner factor~\cite{Mazin1,Mazin2} and
the AF instability due to the nesting effect~\cite{Mazin2}, will
compete.

The RuO$_6$ rotation couples mostly with the $d_{xy}$ orbital but not
with the $d_{yz}$, $d_{zx}$ orbitals because the $pd \pi$ type
hybridization between the O(1)-$2p$ and the $d_{xy}$ states will be
significantly reduced by the RuO$_6$ rotation, but those between the
O-$2p$ and the $d_{yz}$, $d_{zx}$ states are not affected so much. The
direct results of this reduced $pd \pi$ type hybridization between the
O(1)-$2p$ and the $d_{xy}$ states are, first the narrowing of $d_{xy}$
band width and second the downward shift of $d_{xy}$ band, as shown in
Fig. 2(b) (about 0.4 eV narrowing and 0.1 eV downward shift of
$d_{xy}$ band for $\phi=12^{\circ}, \lambda=1.07$).  As the latter
brings the VHS closer to the Fermi level, both of the two results will
enhance the DOS at the Fermi level.  Another effect coming from the
downward shift of the $d_{xy}$ states is the population reduction in
the $d_{yz}$, $d_{zx}$ states, which may shift the Fermi surface
nesting vector closer to the zone boundary.  However, the increase of
the DOS at the Fermi level is the dominant effect and the tendency
towards FM instability is enhanced by the RuO$_6$ rotation.  Once
tilting is additionally introduced, all of the $t_{2g}$ bands will
become narrower.  This will enhance the nesting effect and enhance the
AF instability.  The discussion so far can answer the first question.

Now let us discuss the effects of RuO$_6$ flattening.  There are two
factors also.  First, with the flattening of RuO$_6$ octahedron, the
Ru-O(1) bond length will increase, while the Ru-O(1)-Ru angle remains
$180^{\circ}$. The increased bond length will reduce all the $pd \pi$
type hybridizations between the O(1)-$2p$ and the $d_{xy}$, $d_{yz}$,
$d_{zx}$ states.  Therefore, width of all the bands of three
Ru-$t_{2g}$ states is reduced as shown in Fig.2(c) (about 0.4 eV for
$d_{yz,zx}$ bands and 0.3 eV for $d_{xy}$ band for $\lambda=0.96$),
making the DOS at the Fermi level higher. This will favor the FM
solution. Another very important results of flattening is the orbital
polarization. It is obvious that the tetragonal distortion by the
flattening will populate the $d_{xy}$ states and depopulate the
$d_{yz}$, $d_{zx}$ states (about 0.2 eV downward shift of $d_{xy}$
band for $\lambda=0.96$).  The effect is similar to the RuO$_{6}$
rotation.  The orbital polarization due to flattening will also shift
the nesting vector to the zone boundary as shown in the susceptibility
calculations (Fig.3). This will favor the commensurate AF state of the
system.  The net effect by rotation, tilting and flattening of the
RuO$_6$ will depend on the competition among them and the phase
diagram of Fig.1 demonstrates the situation in a space spanned by
those distortion modes.  Figure 2(f) shows the calculated DOS for the
AF state with $\phi=\theta=12^\circ$, i.e. almost the experimental
structure of Ca$_2$RuO$_4$. It is clear in this case that the occupied
minority spin states mostly come from the $d_{xy}$ orbital due to
flattening of RuO$_6$. Therefore, the strong superexchange interaction
between the occupied majority-spin and unoccupied minority-spin
$d_{yz}$, $d_{zx}$ orbitals will stabilize the AF ground
state~\cite{comment2}.

In summary, by constructing a phase diagram of Sr$_2$RuO$_4$ with
structural distortions, we find the strong coupling between the
lattice and the magnetism. Our phase diagram can qualitatively explain
the experimental phase diagram of Ca$_{2-x}$Sr$_x$RuO$_4$.  We
demonstrate that the RuO$_6$ rotation will enhance the FM instability
in the system, while the tilting plus the flattening of RuO$_6$ make
the system AF. We pointed out that the flattening of RuO$_6$ is so
important not only for the AF state but also for the FM state. An
important implication of our results is that the magnetic and the
structural instabilities in Sr$_2$RuO$_4$ should be strongly
correlated. The structure fluctuation and the magnetic fluctuation
cooperate. Actually the phonon mode corresponding to the RuO$_6$
rotation is quite soft~\cite{phonon} in the bulk, and this rotation
will enhance the FM instability. All these results imply the necessity
of reconsidering the coupling mechanism for the unconventional
superconductivity.  In this context, we propose a possible way to
identify experimentally the relationship between the FM fluctuation
and the superconducting state.  As the uniaxial compression of
Sr$_2$RuO$_4$ will enhance the FM fluctuation without introducing the
disorder, the variation of superconducting transition temperature
against uniaxial compression may provide important information.

The authors thank Dr. S. Nakatsuji, Prof. E. W. Plummer,
Prof. Y. Tokura, Dr. R. Matzdorf, and Dr. J. D. Zhang for valuable
discussions and for providing us with their experimental data. One of
the anthors (Z. Fang) also thanks Dr. David Singh for valuable
discussions. The present work is partly supported by NEDO.

\begin{figure}
    \centering
    \epsfig{file=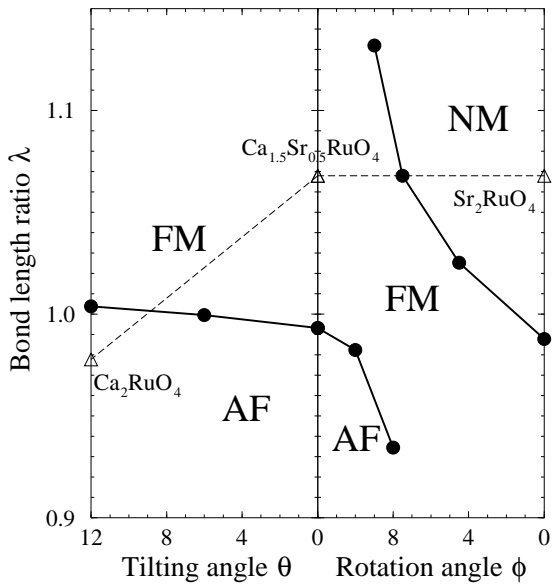,width=75mm}
     \caption{The calculated magnetic phase diagram of Sr$_2$RuO$_4$
       with structural distortions. When the tilting of RuO$_6$
       octahedron is conducted, 12 degrees of RuO$_6$ rotation is
       reserved (see the text for detailed description). The solid
       bold lines are calculated phase boundaries, while the triangles
       linked by dashed line correspond to experimental data. }
\end{figure}

\begin{figure}
    \centering \psfull
    \epsfig{file=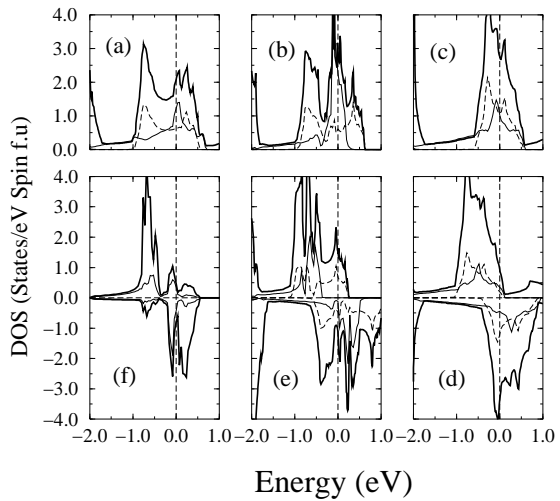,width=75mm}
     \caption{The calculated electronic densities of states (DOS) for
       some particular points in our phase diagram, i.e. (a)
       $\phi$=$\theta$=$0^\circ$, $\lambda=1.07$, NM state;
       (b)$\phi$=$12^\circ$, $\theta=0^\circ$, $\lambda=1.07$, NM
       state; (c)$\phi$=$\theta$=$0^\circ$, $\lambda=0.96$, NM state;
       (d)$\phi$=$\theta$=$0^\circ$, $\lambda=0.96$, FM state;
       (e)$\phi$=$12^\circ$, $\theta$=$0^\circ$, $\lambda=1.07$, FM
       state; (f)$\phi$=$12^\circ$, $\theta$=$12^\circ$,
       $\lambda=0.96$, AF state. The bold solid lines show the total
       DOS (in (f), the local DOS is shown), while the thin solid and
       dashed lines give the projected DOS for the $d_{xy}$ and
       $d_{yx}$, $d_{zx}$ orbitals respectively. Only the regions (-2 eV
       $\sim$ 1 eV), where Ru-$t_{2g}$ states dominate, are shown. The
       Fermi levels are located at the energy zero. }
\end{figure}

\begin{figure}
   \centering
      \leavevmode  \epsfxsize=75mm \epsfbox[50 300 650 800]{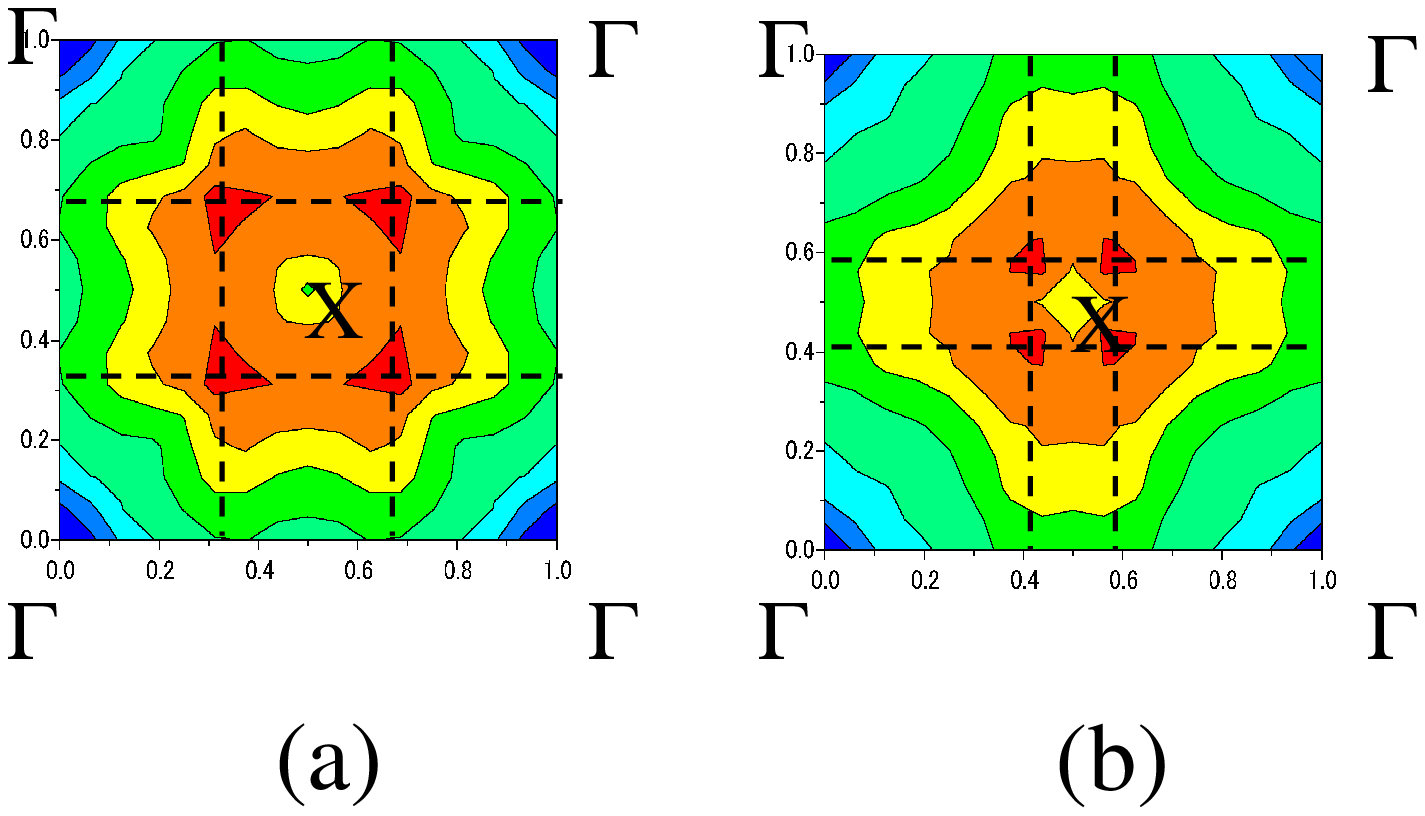}
   \caption{A contour plot of the calculated bare spin susceptibility for
     (a)$\phi$=$\theta$=$0^\circ$, $\lambda=1.07$;
     (b)$\phi$=$\theta$=$0^\circ$, $\lambda=0.96$. The dashed lines
     are guide to the eye for the nesting vectors. The red color
     denote the higher intensity.}
\end{figure}

\newpage
\begin{table}
\caption{The calculated total energies and magnetic moments
   for some particular points in our phase diagram.}
\begin{tabular}{c|c|c|c}
     &NM  &FM &AF  \\  \hline
$\lambda=0.96$    &0 meV   &-40 meV   &-8 meV \\
$\phi=\theta=0^\circ$   &    &1.26 $\mu_B$/Ru   &0.7 $\mu_B$/Ru  \\
\hline
$\lambda=1.07$    &0 meV   &-25 meV   &   \\
$\phi=12^\circ$, $\theta=0^\circ$   &    &0.74 $\mu_B$/Ru   &  \\
\hline
$\lambda=0.96$    &0 meV   &-100 meV   &-117 meV \\
$\phi=12^\circ$, $\theta=12^\circ$   &    &1.13 $\mu_B$/Ru   &0.93
$\mu_B$/Ru  \\
\end{tabular}
\end{table}


\begin{thebibliography}{50}
\bibitem{nature} Y.~Maeno, H.~Hashimoto, K.~Yoshida, {\it et al.},
   Nature (London) {\bf 372}, 532 (1994).

\bibitem{NMR} T.~Imai, A.~W.~Hunt, K.~R.~Thurber and F.~C.~Chou, Phys.
   Rev. Lett. {\bf 81}, 3006 (1998).

\bibitem{Impurity} A.~P.~Mackenzie {\it et al.}, Phys. Rev. Lett. {\bf
     80}, 161 and 3890(1998).

\bibitem{time_reversal} G. M. Luke, Y. Fudamoto, K. M. Kojima and {\it
     et al.}, Nature {\bf 394}, 558(1998).

\bibitem{flux} T. M. Riseman, P. G. Kealey, E. M. Forgan and {\it et
     al.}, Nature {\bf 396}, 242(1998).

\bibitem{Andreev} F.~Laube, G.~Goll, H.~v.~L{\"o}hneysen and {\it et
     al.}, Phys. Rev. Lett. {\bf 84}, 1595(2000).

\bibitem{p-wave} T.~M.~Rice and M.~Sigrist, J. Phys. Condens. Matter
   {\bf 7}, L643 (1995).

\bibitem{Mazin1} I.~I.~Mazin and D.~J.~Singh, Phys. Rev. Lett. {\bf
     79}, 733 (1997).

\bibitem{Tewordt} L.~Tewordt, Phys. Rev. Lett. {\bf 83}, 1007 (1999).

\bibitem{p-wave2} K.~Ishida, H.~Mukuda, Y.~Kitaoka and {\it et al.},
   Nature {\bf 396}, 658(1998).

\bibitem{Sidis} Y.~Sidis, M. Braden, P.~Bourges and {\it et al.},
   Phys. Rev. Lett. {\bf 83}, 3320 (1999).

\bibitem{Mazin2} I.~I.~Mazin and D.~J.~Singh, Phys. Rev. Lett. {\bf
     82}, 4324 (1999).

\bibitem{phonon} M.~Braden and W.~Reichardt, Phys. Rev. B {\bf 57},
   1236(1998).

\bibitem{plummer} R.~Matzdorf, Z. Fang, X. Ismail and {\it et al.},
   Science {\bf 289}, 746 (2000).

 \bibitem{Nakatsuji} S. Nakatsuji and Y. Maeno, Phys. Rev. Lett. {\bf
     84}, 2666(2000); {\it ibid}, Phys. Rev. B {\bf 62}, 6458 (2000).

\bibitem{Friedt} O. Friedt, M. Braden, G. Andr\'{e} and {\it et al.},
   cond-mat/0007218.

 \bibitem{Mazin3} I.~I.~Mazin and D.~J.~Singh, Phys. Rev. B {\bf 56},
   2556 (1997).

\bibitem{PP} D. Vanderbilt, Phys. Rev. B {\bf 41}, 7892 (1990).

\bibitem{NC} N.~Troullier and J.~L.~Martins, Phys. Rev. B {\bf 43},
   1993(1991).

\bibitem{CRO} M. Braden, G. Andr\'{e}, S. Nakatsuji and Y. Maeno,
   Phys. Rev. B {\bf 58}, 847 (1998).

 \bibitem{comment1} The experimental information [16] was adopted for
   the structural changes in the phase diagram. However, the real
   alloy system Ca$_{2-x}$Sr$_x$RuO$_4$ can be treated by the virtual
   crystal approximation (VCA) [Z. Fang {\it et al.}, Phys. Rev. Lett.
   {\bf 84}, 3169 (2000)]. Our test calculations in the LDA for
   Ca$_{1.5}$Sr$_{0.5}$RuO$_4$ with experimental lattice parameters
   give the FM ground state with the optimized RuO$_6$ rotation angle
   $\phi$=9.5$^{\circ}$ and bond length ratio $\lambda$=1.076.

 \bibitem{comment2} For Ca$_2$RuO$_4$, the generalized gradient
   approximation (GGA) may be better than LDA due to the narrowing of
   bands. However, our test calculations for Ca$_2$RuO$_4$
   ($\phi=\theta=12^{\circ}, \lambda=0.96$) suggested that the total
   energy difference between the AF and the FM solutions changes only
   slightly (from -17 meV in LDA to -20 meV in GGA), although a tiny
   gap can be obtained in GGA.






\end{thebibliography}
\end{document}